\documentclass{aa}  

\usepackage{graphicx}
\usepackage{txfonts}
\usepackage[colorlinks = true, allcolors=blue]{hyperref}
\usepackage[normalem]{ulem}
\usepackage{booktabs}
\usepackage{orcidlink}
\newcommand\arcdeg{\mbox{$^\circ$}}%
\newcommand\farcdeg{\mbox{$.\!\!^\circ$}}%
\newcommand{\Msun}{\hbox{M$_\odot$}}
\begin{document}

   \title{UVIT Open Cluster Study. X.}
   \subtitle{Rich Collection of Post Mass Transfer Systems in NGC 6791}

   \author{Vikrant V. Jadhav\orcidlink{0000-0002-8672-3300}\inst{1,2},
          Annapurni Subramaniam\orcidlink{0000-0003-4612-620X}\inst{3}
          \and
          Ram Sagar\orcidlink{0000-0003-4973-4745}\inst{3}
          }

   \institute{
        Helmholtz-Institut für Strahlen- und Kernphysik, Universität Bonn, Nussallee 14-16, D-53115 Bonn, Germany\\
        \email{vjadhav@astro.uni-bonn.de}
        \and
        Inter-University Centre for Astronomy and Astrophysics (IUCAA), Post Bag 4, Ganeshkhind, Pune 411007, India
        \and
        Indian Institute of Astrophysics, Koramangala II Block, Bangalore-560034, India\\
        \email{purni@iiap.res.in, ramsagar@iiap.res.in}
        }

   \date{Received January 13, 2023; accepted June 27, 2023}

 
  \abstract
   {}
   {NGC 6791 is one of the richest old open clusters in the Milky Way. Its position above the Galactic plane and the number density makes it an interesting middle ground between Galactic open and globular clusters. We aim to detect the UV bright population of NGC 6791 using \textit{AstroSat}/UVIT images in near-UV and far-UV filters and characterise the known post mass transfer systems such as blue straggler stars (BSSs).
   }
   {
   We identified 20 members with large UV flux (out of 91 cluster members among 1180 detections), suggestive of binarity, interactions or stellar activity using multi-wavelength spectral energy distribution analysis.}
   {We characterised 62 isolated cluster members, including five hot subdwarfs (sdA/sdB). Additionally, we detected ten sdA/sdB/extremely low mass white dwarf (ELM) type candidates hidden alongside other cluster members.
   Additionally, we report the discovery of four candidate blue lurkers, which are main sequence stars with mass accretion history.
   }
   {We report that this cluster has a variety of stellar (pre-)remnants, such as sdBs, sdAs, and ELM white dwarfs, that are by-products of binary evolution.
   The above are likely to be post mass transfer binaries found throughout the evolutionary phases from the main sequence to the post horizontal branch. Therefore, this dynamically old open cluster is unique, making it an ideal testbed for dynamical studies.
   }

   \keywords{(Galaxy:) Open clusters and associations: individual: NGC 6791 -- (stars:) binaries: general --
                Ultraviolet: stars --
                Catalogues
               }

   \maketitle
%

\section{Introduction}
Open clusters are integral to studying stellar populations due to their relative homogeneity (in metallicity and age) and robust cluster membership due to precise astrometry. UV imaging has proved very useful in identifying optically subluminous stars such as white dwarfs (WDs) and hot subdwarfs \citep{Sahu2019ApJ...876...34S, Jadhav2021JApA...42...89J, Rao2022MNRAS.516.2444R}.
We are currently conducting UVIT Open Cluster Study (UOCS) to understand the UV bright population, focusing on post mass transfer systems. M67 has proven to be home to multiple extremely low mass (ELM) WDs, which are products of mass transfer (\citealt{Sindhu2019ApJ...882...43S, Jadhav2019ApJ...886...13J, Subramaniam2020JApA...41...45S, Pandey2021MNRAS.507.2373P}; Vernekar et al. 2023, in press). Similar mass transfer binaries have been found in NGC 7789 \citep{Vaidya2022MNRAS.511.2274V} and NGC 2506 \citep{ Panthi2022MNRAS.516.5318P}. Here, we extend this study to the open cluster NGC 6791 ($\alpha_{2000}$ = 19h 20m 53s; $\delta_{2000}$ = +37\arcdeg 46\arcmin 18\arcsec; l = 69\farcdeg959; b = +10\farcdeg904) which is $\sim$ 8.5 Gyr old and also one of the metal-rich cluster ([Fe/H] $\sim$ 0.4) known in the Milky Way \citep{Bossini_2019A&A...623A.108B}. It is located at a distance of $\sim$ 4.1 Kpc and is well studied massive ($\sim5000$ \Msun) open star cluster.  

Using ground-based and \textit{Kepler} photometry and multi-epoch spectroscopy data, \citet{Brogaard2018MNRAS.481.5062B} identified the binary star V106 as a blue straggler star (BSS) member of the NGC 6791. They derived the primary mass as 1.67 \Msun, more massive than the cluster turn-off mass and the secondary star as a bloated (proto) ELM helium WD. A detailed study also reveals that V106 is potentially a prototype progenitor of old field giants masquerading as young. \citet{Villanova2018ApJ...867...34V} presented and discussed detailed abundances of 17 evolved stars of NGC 6791 using high-resolution spectra obtained with the UltraViolet and Visible Echelle Spectrograph at the European Southern Observatory Very Large Telescope and High Resolution Echelle Spectrometer at the Keck telescope. They obtained a mean [Fe/H] = $+0.313 \pm 0.005$, in good agreement with recent estimates. \citet{Tofflemire2014AJ....148...61T} provided the epoch radial velocity (RV) and related results for 280 stars, including main sequence (MS), red giant branch (RGB) and horizontal branch (HB) stars.
\citet{Jadhav2021MNRAS.507.1699J} identified 47 potential BSSs in NGC 6791 using \textit{Gaia} DR2 data. 
\citet{Kamann2019MNRAS.483.2197K} combined \textit{Gaia} data with the archival line of sight velocities and studied the internal dynamics of the NGC 6791 in three dimensions. 
\citet{Martinez2018MNRAS.474...32M} performed an orbital analysis within a Galactic model (including spiral arms and a bar) and found that it is plausible that NGC 6791 formed in the inner thin disc or the bulge and was later displaced by radial migration to its current orbit. The birthplace and journeys of NGC 6791 are imprinted in its chemical composition, mass-loss and flat stellar mass function, supporting its origin in the inner thin disc or the bulge. 

NGC 6791 has one of the largest populations of BSSs among all known open clusters \citep{Jadhav2021MNRAS.507.1699J}. The BSSs result from binary interactions such as collisions \citep{Hills1976ApL....17...87H} or mergers \citep{Perets_2009ApJ...697.1048P}. However, knowing which formation pathway had been followed is difficult due to the incomplete parameterisation of such systems. The mass of the BSS, stellar rotation, abundance peculiarities and characteristics of the donor remnant can be used to identify the formation pathway.

In this work, we analysed the UV bright population of NGC 6791 using multi-wavelength spectral energy distributions (SEDs). 
\S \ref{sec:obs} gives the details of the data and analytical methods, \S \ref{sec:results} presents the results and their implications.

\begin{table}[!hb]
    \caption{Exposure times in UVIT filters and detected sources.}
    \label{tab:UVIT_data}
    \centering
    \begin{tabular}{cccc}
    \toprule
        Filter	&	Exp. Time [s]	&	Detections	&	Members	\\ \toprule
        F172M	&	3017.251	&	55	&	6	\\
        N245M	&	1259.539	&	734	&	35	\\
        N263M	&	1156.669	&	1180	&	81	\\
        N279N	&	364.296	&	165	&	8	\\ \bottomrule
    \end{tabular}
\end{table}

\section{Data and analysis} \label{sec:obs}

\subsection{UVIT data and \textit{Gaia} membership}

We observed NGC 6791 in August 2017 with the Ultra-Violet Imaging Telescope (UVIT) onboard \textit{AstroSat} observatory (proposal ID: A03\_008). The UVIT performs observations in far-UV (130--180 nm), near-UV (NUV; 200--300 nm) and VIS (350--550 nm) channels. More details of the calibration and instrumentation are present in \citet{Tandon2017AJ....154..128T,Tandon2020AJ....159..158T,Kumar2012SPIE.8443E..1NK}. The exposure times, source detections and other observational details are given in Table~\ref{tab:UVIT_data}. 
The UVIT data reduction and astrometry were performed using \textsc{ccdlab} \citep{Postma2017PASP..129k5002P, Postma2020PASP..132e4503P}.
The point spread function photometry of the UVIT images was done with \textsc{iraf} \citep{Tody1993ASPC...52..173T}. 
Preliminary analysis and cross-matching were done with \textsc{topcat} \citep{Taylor2005ASPC..347...29T}.

\begin{figure*}[!ht]
    \centering
    \includegraphics[width=0.95\textwidth]{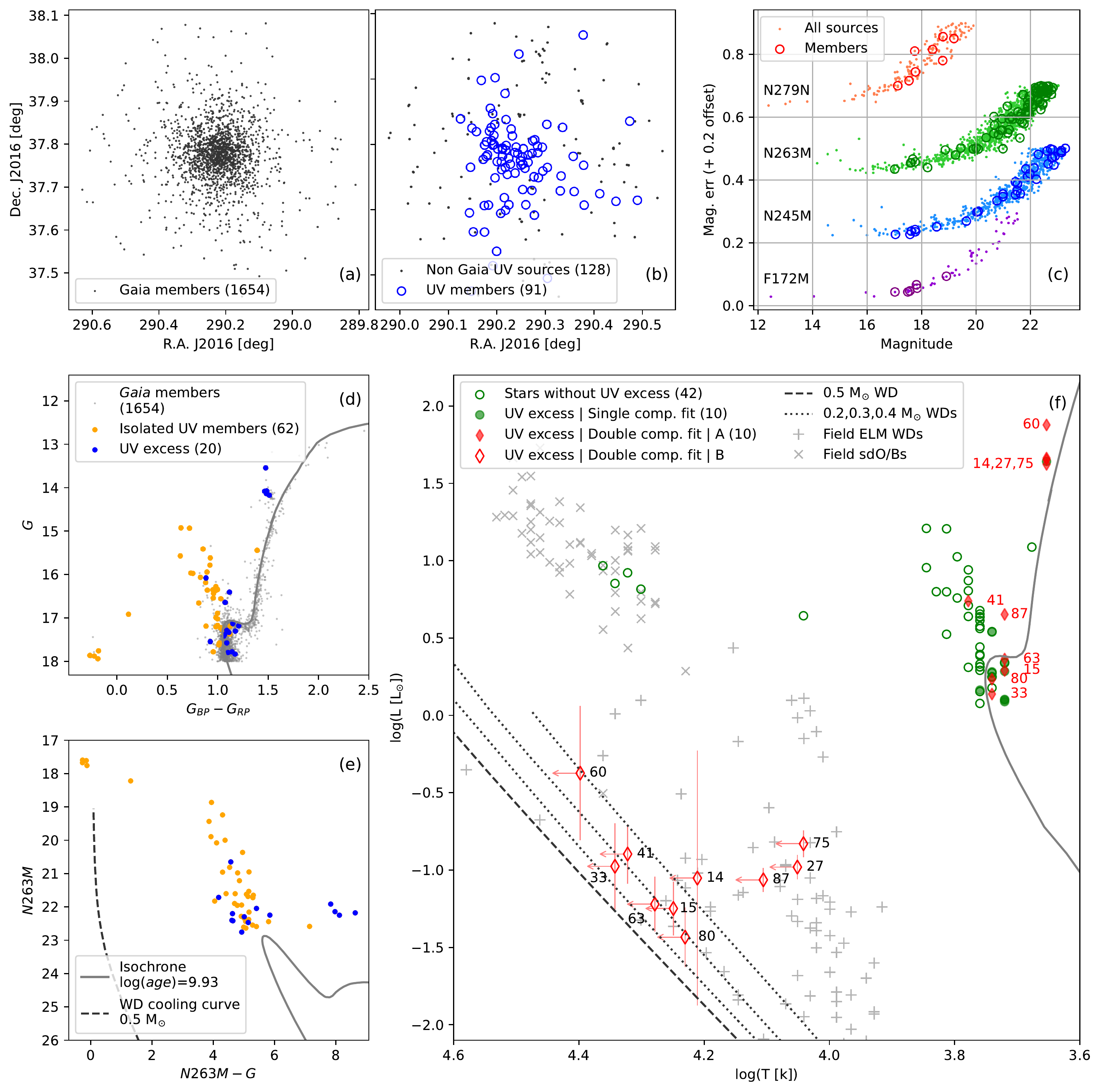}
    \caption{(a) Spatial distribution of \textit{Gaia} members. (b) Spatial distribution of UV detected members (blue circles) and UV detected sources which are not present in \textit{Gaia} EDR3 catalogue (grey dots). (c) Magnitude vs error plots for UVIT photometry.
    (d) \textit{Gaia} CMD showing the \citeauthor{Cantat2020A&A...640A...1C} members (as grey), UV detected members (as orange) and members with UV excess (as blue). (e) The UV-optical CMD of cluster members. (f) The HR diagram created using SED fitting results. The figure shows 42 sources without UV excess (hollow green circles) and ten sources with UV excess (green filled circles) modelled using single-component SED. The ten sources with double-component fitting are shown as hollow red squares and red-filled squares for A and B components, respectively.
    Their corresponding IDs are also shown. The isochrone of log($age$) = 9.93 (grey curve), 0.5 \Msun\ WD cooling curve (black dashed curve) and 0.2--0.4 \Msun\ WD cooling curves (black dotted curves) are shown wherever needed in (d)--(f).}
    \label{fig:spatial}
\end{figure*}

We have used \textit{Gaia} DR2 \citep{Gaia2016A&A...595A...1G, Gaia2018A&A...616A...1G} based membership catalogue by \citet{Cantat2020A&A...640A...1C}. It is one of the latest membership catalogues available, and the inclusion of \textit{Gaia} DR3 data does not significantly improve the membership for the bright stars targeted in this study.
The catalogue contains 1654 members with probability $\geq0.6$. Fig.~\ref{fig:spatial} (a) shows the spatial distribution of the \textit{Gaia} members. 
Fig.~\ref{fig:spatial} (b) shows the spatial distribution of UVIT-detected sources.
The magnitude vs error distribution for UVIT filters is shown in Fig.~\ref{fig:spatial} (c). 
Among the \textit{Gaia} members, 91 are detected in at least one UV filter. 
The numbers of UV-detected members in individual filters are given in Table~\ref{tab:UVIT_data}.

\subsection{Colour magnitude diagrams}
Fig.~\ref{fig:spatial} (d) shows the \textit{Gaia} colour-magnitude diagrams (CMDs) of NGC 6791. The isochrones shown are generated using PARSEC v3.7 \citep{Bressan2012MNRAS.427..127B} with log($age$) = 9.93, [M/H]=0.4, distance modulus = 13.085 mag and A$_v$ = 0.302 mag \citep{Bossini_2019A&A...623A.108B}. Also shown are the WD cooling curves (\citealt{Bedard2020ApJ...901...93B} and references therein) with a mass of 0.5 \Msun, which is the expected mass of WDs based on the cluster turn-off and WD initial-final mass relation \citep{Cummings2018ApJ...866...21C}.
Fig.~\ref{fig:spatial} (d) shows that UVIT has detected stars in the MS, RGB, BSS and hot subdwarf (sdA, sdB) phases. The sources with UV excess (see \S\ref{sec:SEDs}) are present on the MS and RGB. The UV-optical CMD in Fig.~\ref{fig:spatial} (e) shows the UV-detected members in UVIT/N263M filter.

\subsection{Spectral energy distributions}
\label{sec:SEDs}

\begin{figure*}[!ht]
    \centering
    \includegraphics[width=0.95\textwidth]{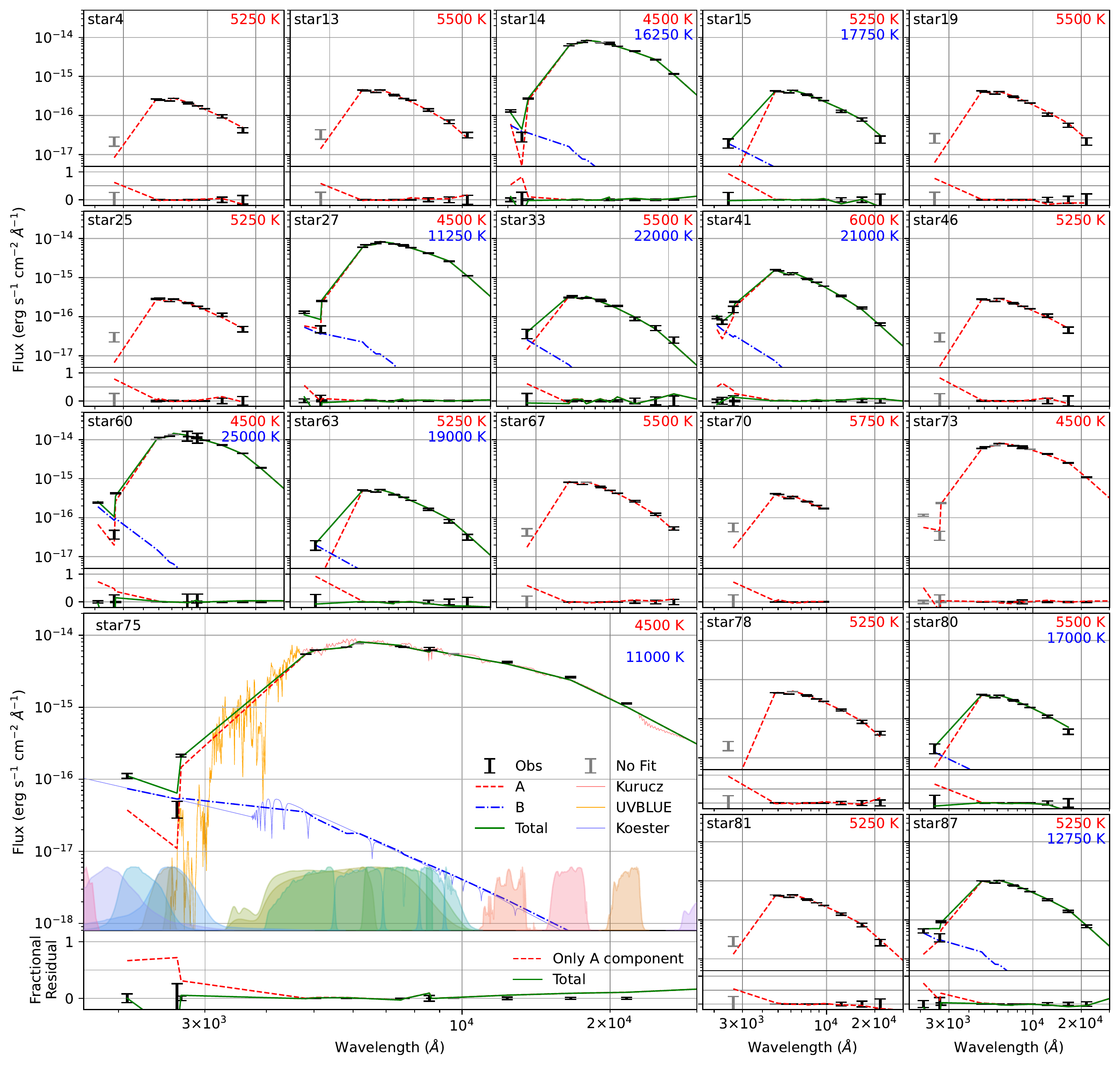}
    \caption{SEDs of stars with fractional UV excess of more than 0.5 in at least one filter. The top panel for each star shows the observed flux (black error bars), data omitted while fitting SED (grey error bars), fitted Kurucz+UVBLUE model (red dashed curve), fitted Koester model (blue dot-dashed curve; wherever applicable) and total model flux (green curve; wherever applicable). The bottom panel shows the residual flux after the single component fit (red dashed curve) and double component fitting (green curve). The panel for star75 shows extra details such as the high-resolution spectral models (Kurucz: red, UVBLUE: orange, Koester: blue) and the used filter transmission curves (filled curves at the bottom).
    }
    \label{fig:SEDs_1}
\end{figure*}

We visually checked the neighbourhoods ($\sim$3\arcsec) 
of all 91 UV-detected sources to remove 29 sources with bright neighbours. We then constructed the SEDs of 62 isolated sources using \textit{AstroSat}/UVIT, \textit{Gaia} EDR3 \citep{Gaia2021A&A...649A...1G}, \textit{swift}/UVOT \citep{Siegel2019AJ....158...35S}, Pan-STARRS \citep{Chambers2016arXiv161205560C}, 2MASS \citep{Skrutskie2006AJ....131.1163S} and \textit{WISE} \citep{Wright2010AJ....140.1868W}. The virtual observatory cross-matching was done using \textsc{vosa} \citep{Bayo2008A&A...492..277B} and Vizier \citep{Ochsenbein2000A&AS..143...23O}. The sources were extinction corrected using known extinction--wavelength relations \citep{Fitzpatrick1999PASP..111...63F, Indebetouw2005ApJ...619..931I}. 

We used \textsc{vosa}\footnote{\url{http://svo2.cab.inta-csic.es/theory/vosa/index.php}} to fit Kurucz models \citep{Castelli2003IAUS..210P.A20C} to the SEDs using $\chi^2$ minimisation technique. The metallicity of the Kurucz model was fixed to +0.5 (closest to the cluster metallicity of +0.4). Also, we fixed the distance to 4139$\pm$100 pc and A$_v$ to 0.302$\pm$0.005. The free parameters were temperature (3500 to 50000 K) and log$g$ (2 to 5). However, note that the SED-derived log$g$ parameter is imprecise.
After fitting the SEDs, we calculated the fractional residual as follows:
\begin{equation}
    Fractional\ residual = \frac{F_{obs}-F_{model}}{F_{obs}}
\end{equation}
The fractional residual is shown in the bottom sub-panels of the SEDs (Fig.~\ref{fig:SEDs_1}).

Twenty-nine sources showed more than 0.5 fractional residual in at least one UV filter. 
In the UOCS series of papers, we have been using Kurucz models to fit the SEDs, as they cover a wide range of wavelengths. In this study, we checked whether the detected UV excess changes are due to underestimating the Wein tail flux in the Kurucz models. Hence, we created a hybrid Kurucz+UVBLUE model which uses specialised UVBLUE models \citep{Rodriguez2005ApJ...626..411R} in 850--4700 \AA\ region and Kurucz models for the remaining red part of the spectrum. The UVBLUE models were convolved with the UVIT and UVOT filters (where the complete filter falls within UVBLUE's range) to get synthetic photometry for a temperature range of 3000 to 50000 K, log$g$ range of 0 to 5 and metallicity range of $-2$ to 0.5. The filter transmission profiles were taken from SVO Filter Profile Service\footnote{\url{http://svo2.cab.inta-csic.es/theory/fps/}} \citep{Rodrigo2012ivoa.rept.1015R, Rodrigo2020sea..confE.182R}.
We could fit nine stars as single stars with this hybrid model using the python code \textsc{Binary\_SED\_Fitting}\footnote{\url{https://github.com/jikrant3/Binary\_SED\_Fitting}} \citep{Jadhav2021JApA...42...89J}. This leaves 20 sources with UV excess using the Kurucz+UVBLUE models.

The UV excess has been linked to the presence of hot compact objects such as WDs \citep{Jadhav2019ApJ...886...13J, Rao2022MNRAS.516.2444R, Panthi2022MNRAS.516.5318P} or hot subdwarfs \citep{Jadhav2021JApA...42...89J}. One can use double-component SED fitting to deconvolve the two stars.
As \textsc{vosa} cannot use hybrid models, we used \textsc{Binary\_SED\_Fitting} to perform double component fits using the hybrid model for the primary star (Kurucz+UVBLUE) and Koester model \citep{Tremblay2009ApJ...696.1755T, Koester2010MmSAI..81..921K} for the possible hotter companion (see \citealt{Jadhav2022arXiv220703780J} for more details). Ten sources could be fitted with satisfactory double components, while the remaining ten could not be fitted with available data.

The results of single and double-component SED fitting are given in Table~\ref{tab:results_short}. An extended table, which includes photometry and fitting parameters, is available as supplementary material and on CDS. Fig.~\ref{fig:SEDs_1} shows the SED fits for single and double-component fitting for the 20 sources with UV excess. 
The observed data points and associated error bars are shown in black colour. 
The line plots connecting estimated flux from synthetic photometry of the various models are shown in different colours 
(red: Kurucz+UVBLUE model, blue: Koester model, green: total flux of Kurucz+UVBLUE and Koester models).
In the double SEDs, the cooler component (Kurucz or Kurucz+UVBLUE model) is denoted as `A', while the hotter component (Koester model) is denoted as `B'.

In the double-component fitting, the parameters for the cooler model SED are improved as the effects of any potential hotter companion or unusual activity are reduced.
However, we have found that the parameter estimates derived from only 1--2 UV data points are only partially reliable. The luminosity values are relatively accurate, but the temperature may be underestimated ($\equiv$ overestimated radius; Jadhav et al. in prep). Hence, we recommend not using the temperature/radius estimates of the hotter model SEDs.

\section{Results} \label{sec:results}

\begin{table*}
    \caption{The list of target stars along with the comments from \citet{Tofflemire2014AJ....148...61T, Sanjayan2022AcA....72...77S} and this work. 
    BSS: Blue straggler stars, 
    ECL: Eclipsing Binary, 
    HB: horizontal branch,
    Puls: pulsator, 
    RC: red clump,  
    Rot. var: rotational variable, 
    RR: rapid rotator.
    An extended table with UVIT photometry for all 91 UV detected members and SED fitting parameters of 62 isolated sources is given as supplementary material and available at the CDS.}
    \label{tab:results_short}
    \centering
    \begin{tabular}{lrrc|lrrc}
\toprule
Name & $\alpha_{J2016}$ [$^{\circ}$] & $\delta_{J2016}$ [$^{\circ}$] & Comment & Name & $\alpha_{J2016}$ [$^{\circ}$] & $\delta_{J2016}$ [$^{\circ}$] & Comment \\
\midrule
\multicolumn{3}{r}{No	UV	excess} &                                  & star56 &  290.257766 &  37.763117 &                                 MS \\ 
star0  &  290.182556 &  37.706893 &                                    & star57 &  290.178641 &  37.803567 &                                BSS \\
star1  &  290.330379 &  37.728380 &                                    & star58 &  290.377713 &  37.966663 &                                BSS \\
star2  &  290.272239 &  37.785515 &                                    & star59 &  290.196283 &  37.748383 &                                    \\
star3  &  290.278053 &  37.806657 &                                    & star61 &  290.273850 &  37.804970 &                                    \\
star5  &  290.180772 &  37.783681 &                                    & star62 &  290.302801 &  37.830190 &                                BSS \\
star6  &  290.143487 &  37.695005 &                                    & star64 &  290.411506 &  37.724085 &                                 MS \\
star7  &  290.263954 &  37.783218 &                      sdB, Puls & star65 &  290.217391 &  37.764303 &                                    \\
star8  &  290.174187 &  37.706013 &                                 MS & star66 &  290.215274 &  37.718092 &                                BSS \\
star9  &  290.229360 &  37.822990 &                                    & star68 &  290.211291 &  37.792004 &                                    \\
star10 &  290.167968 &  37.897422 &                                sdA & star69 &  290.248584 &  37.774346 &       BSS, RR, Variable \\
star11 &  290.237421 &  37.755679 &                                BSS & star71 &  290.186347 &  37.854764 &                                 MS \\
star12 &  290.332132 &  37.751261 &                                 MS & star72 &  290.303946 &  37.594908 &                                 MS \\
star16 &  290.200673 &  37.773090 &                                    & star74 &  290.209131 &  37.772973 &                                BSS \\
star17 &  290.293412 &  37.795104 &            BSS, SB1, RR & star76 &  290.145732 &  37.764532 &  BSS, SB1, RR, Variable \\
star18 &  290.191630 &  37.798388 &                                    & star77 &  290.147393 &  37.575227 &                                BSS \\
star20 &  290.217878 &  37.698901 &                                 MS & star79 &  290.188213 &  37.825343 &                      sdB, Puls \\
star21 &  290.242940 &  37.785566 &                           BSS, SB1 & star82 &  290.231777 &  37.812650 &                                    \\
star22 &  290.224330 &  37.778367 &                                    & star83 &  290.264080 &  37.772164 &                                    \\
star23 &  290.254358 &  37.792965 &                                    & star84 &  290.253508 &  37.769471 &                                BSS \\
star24 &  290.191568 &  37.614156 &                                BSS & star85 &  290.357116 &  37.781353 &                                 MS \\
star26 &  290.259442 &  37.707294 &                                 MS & star86 &  290.253143 &  37.730550 &                                    \\
star28 &  290.245497 &  37.937647 &                                BSS & star88 &  290.358896 &  37.727381 &                                 MS \\
star29 &  290.201645 &  37.792628 &                                    & star89 &  290.211412 &  37.793906 &                                    \\
star30 &  290.194181 &  37.799349 &                                BSS & star90 &  290.249771 &  37.767570 &                                BSS \\
star31 &  290.221496 &  37.753805 &                                    & \multicolumn{4}{c}{UV	excess	-	Single	fit}                        \\
star32 &  290.284253 &  37.786966 &                                    & star4  &  290.196492 &  37.901509 &                                 MS \\
star34 &  290.236587 &  37.787253 &                                 MS & star13 &  290.239528 &  37.795310 &                                 MS \\
star35 &  290.199415 &  37.636157 &                                 MS & star19 &  290.195735 &  37.830140 &                                 MS \\
star36 &  290.216948 &  37.877327 &                                    & star25 &  290.184428 &  37.819743 &                                 MS \\
star37 &  290.276987 &  37.691396 &                                    & star46 &  290.378518 &  37.767541 &                                 MS \\
star38 &  290.473241 &  37.834930 &                                 MS & star67 &  290.252538 &  37.780422 &                                BSS \\
star39 &  290.193870 &  37.813223 &                                    & star70 &  290.271927 &  37.729174 &                                 MS \\
star40 &  290.209807 &  37.760943 &                                BSS & star73 &  290.174902 &  37.798373 &                                 RC \\
star42 &  290.224628 &  37.707220 &                           BSS, SB1 & star78 &  290.148421 &  37.824601 &                                 MS \\
star43 &  290.218488 &  37.790879 &                                    & star81 &  290.444898 &  37.712358 &                                 MS \\
star44 &  290.207915 &  37.694110 &                                    & \multicolumn{4}{c}{UV	excess	-	Double	fit}                        \\
star45 &  290.379465 &  37.706632 &                                 MS & star14 &  290.218634 &  37.837626 &                                 RC \\
star47 &  290.294602 &  37.758718 &            BSS, $\gamma$ Dor Puls & star15 &  290.157917 &  37.819014 &                                 MS \\
star48 &  290.188867 &  37.805342 &                                sdB & star27 &  290.276235 &  37.749896 &                                 RC \\
star49 &  290.303713 &  37.764194 &                      sdB, Puls & star33 &  290.151458 &  37.665758 &               MS, ECL \\
star50 &  290.191585 &  37.846677 &                                RGB & star41 &  290.211737 &  37.781781 &           BSS, Rot. var. \\
star51 &  290.285072 &  37.748605 &                                    & star60 &  290.283629 &  37.797057 &                            post-HB \\
star52 &  290.174360 &  37.665663 &                                 MS & star63 &  290.335191 &  37.825374 &                                 MS \\
star53 &  290.218961 &  37.797860 &           BSS, Rot. var. & star75 &  290.125237 &  37.838649 &                                 RC \\
star54 &  290.489114 &  37.714080 &                                    & star80 &  290.171933 &  37.776700 &                                 MS \\
star55 &  290.177668 &  37.781130 &                 BSS, RR & star87 &  290.227669 &  37.769334 &                                BSS \\
\bottomrule

    \end{tabular}

\end{table*}

Fig.~\ref{fig:spatial} (d) shows that most of the BSSs are detected in NUV along with some MS and RGBs. We found UV excess in 20 of these sources. Among them, 12 are MS stars, three are BSSs, and five are giants. Fig.~\ref{fig:spatial} (e) demonstrates the excess flux in these sources as the MS stars and RGBs are brighter than the isochrone, and in fact, all the sources detected in the UV and located on the MS have some fraction of excess flux in the NUV. The remaining 42 isolated UVIT-detected stars (21 BSSs, 15 MS, one RGB, four sdBs and one sdA) did not show significant UV excess.

The SED fitting gives the temperatures, luminosities, and radii of the sources. The temperature range for BSSs is 5250--7000 K while for MS stars it is 5250--6000 K. Fig.~\ref{fig:spatial} (f) shows the Hertzsprung--Russell 
 (HR) diagram of the SED  fitting results. 
We have also included the parameters of the potential hotter companions. If they are real, their luminosities will be of the order of 0.1 L$_{\odot}$. 
Their SED-based temperatures are 11000--25000 K, and the radii are 0.02--0.11 R$_{\odot}$. They could be (i) sub-luminous sdAs/sdBs or (ii) ELM WDs or (iii) bloated proto-WDs (e.g. \citealt{Brogaard2018MNRAS.481.5062B}) or (iv) their actual temperature is significantly larger, and thus these radii are overestimated which could make them typical WDs.
The grey markers in the background show the HR diagram positions of field ELMs and field sdO/sdBs for comparison. 
The field ELMs are a sub-sample of sources in  \citet{Brown2016ApJ...818..155B}.
The field sdO/sdBs are a sub-sample of hot subdwarfs given in \citet{Geier2020A&A...635A.193G}. 
Both samples were fitted with single-component Kurucz SEDs using \textsc{vosa}.

In the course of the SED fitting, we also tried making only Kurucz and Koester model binary SEDs. That resulted in 29 sources with UV excess, among which ten gave satisfactory double fits. Among these, six were common with the hybrid and Koester fit, while four are currently classified as single fits with UV excess (star4, star19, star70 and star73). Four current binary systems (star15, star63, star80 and star87) could not be fit satisfactorily with Kurucz and Koester models using \textsc{vosa}.
The total number of satisfactory binary fits is the same if we include UVBLUE models. However, there are an equal number of unsatisfactory fits when the UVBLUE models are included or excluded. We explored if there is any specific dependency, and in the case of UVBLUE models, the NUV flux for stars cooler than 6000 K is more than the Kurucz models, whereas the difference is marginal in the case of hotter stars. In general, the estimated temperatures of hotter companions were found to be higher if the UVBLUE models were incorporated.
Overall, the number of sources with successful binary fits is the same in either approach; however, the overall number of sources with UV excess is reduced if UVBLUE is included.

\subsection{Comments on individual systems}

We did not create SEDs of UV-detected stars with neighbours within 5\arcsec. Among them, there is a notable contact binary, 
V5 (P = 0.31265938 d; \citealt{Sanjayan2022AcA....72...77S}; \citealt{Mochejska2003AJ....125.3175M}). 
There are also a few variables (long period, non-periodic, rotational variables): 
KIC 2437079 \citep{Sanjayan2022AcA....72...77S}, 
V65 (P $\sim$ 11.3 d), 
V66 (P $\sim$ 49 or 99 d), and 
V19 \citep{Mochejska2003AJ....125.3175M}. 
These stars are not discussed further due to close neighbours.

\textit{Hot subdwarfs:}
star7 (B5, KIC 2437937), star49 (B4, KIC 2438324), and star79 (B3, KIC 2569576) are known pulsators \citep{Sanjayan2022AcA....72...77S}. 
star48 (B6) and star79 are RV variables and likely binary systems \citep{Sanjayan2022MNRAS.509..763S}. 
star49 is also an RV variable with P = 0.398495 d and a low mass MS companion \citep{Sanjayan2022MNRAS.509..763S}\footnote{\citet{Mochejska2003AJ....125.3175M} stated P = 0.796993 d $\sim2\times0.398495$ d}. The SED temperatures (using Kurucz models) for the three hot subdwarfs are slightly underestimated compared to \citet{Sanjayan2022MNRAS.509..763S} (using Tlusty models), likely due to differences in the models.

\textit{Blue straggler stars:}
star41 (KIC 2437238) showed rotational variability \citep{Sanjayan2022AcA....72...77S} and significant UV excess hinting at a $\gtrapprox0.25$ \Msun\ WD as its companion. The absence of eclipses and RV variability suggests a low inclination orbit in the potential binary system.
The single-lined spectroscopic binary (SB1) and $\gamma$Dor type star47 (V106, KIC 2438249, WOCS 54008) showed no excess UV flux, which is expected as \citet{Brogaard2018MNRAS.481.5062B} derived the temperatures of binary components as 7110 + 6875 K and their masses as 1.62 + 0.176 \Msun. The low temperatures are not enough to create significant UV flux (note that the SED estimated radius of the BSS is overestimated as we modelled both components as a single source). The system is an eclipsing W Uma binary system with P = 1.4464 d \citep{Tofflemire2014AJ....148...61T}. The absence of UV excess and the low temperature of the companion suggests that significant time has passed since the mass transfer event. The system is a prime candidate for a detailed asteroseismic study to constrain the internal structure and mass transfer history in BSSs.
star53 (KIC 2437338) also showed rotational variability \citep{Sanjayan2022AcA....72...77S} but no UV excess or RV variability.
star61 showed UV excess, but we could not fit a hotter companion.
star69 (KIC 2437745) is a rapid rotator and photometric variable (P = 1.43) \citep{Tofflemire2014AJ....148...61T, Sanjayan2022AcA....72...77S}. However, it did not show UV excess. 
star76 (WOCS 46008, KIC 2436421) is SB1 and a rapid rotator \citep{Tofflemire2014AJ....148...61T}. \citet{Sanjayan2022AcA....72...77S} detected variability in its \textit{Kepler} light curves with period of 1.44 days. However, we did not detect any UV excess and could not confirm whether the RV variability is periodic (with 1.44 d) due to poor sampling.
star87 shows UV flux compatible with an ELM WD.
star90 is an RV constant star with no UV excess. \citet{Tofflemire2014AJ....148...61T} suggested that it is a 1.9 \Msun\ merger product formed 1 Gyr ago in line with the absence of any hotter companion. 

\textit{Red clump stars:} There are 19 red clump stars identified using asteroseismic measurements \citep{Stello2011ApJ...739...13S} in NGC 6791. UVIT detected four of them, all having significant UV excess (only three are successfully fitted with a double SED). Among them, star27 (KIC 2438051), star73 (KIC 2436732) and star75 (KIC 2569055) are known pulsating red clump stars.
These red clump stars did not show detectable RV variability or eclipses. 

\textit{Main sequence star:}
star33 (\textit{Gaia} DR3 2051286413614206208) has been classified as eclipsing binary with a period of 0.3664 d \citep{Gaia2022arXiv220800211G, Mowlavi2022arXiv221100929M}. The \textit{Gaia} light curves and the double-component fitting both suggest a close hotter star as its companion. However, more RV variation analysis is needed to confirm the companion and any mass transfer history. The HR diagram positions of companions to star15, star63 and star80 indicate that they are likely WDs with mass $\gtrapprox 0.3$ \Msun.

\textit{Post horizontal branch star:}
star60 (WOCS 9007) is a proper motion member of the cluster. However, its RV is different from the cluster by $\sim25$ km s$^{-1}$ with a parallax of 0.2901$\pm$0.0115 mas yr$^{-1}$ (cluster parallax is 0.241 mas yr$^{-1}$). It showed significant UV excess indicative of a young ELM companion. However, if the star is not a cluster member, then the distance measurement and the derived classification may be incorrect.

\section{Discussion} \label{sec:discusion}
The five single hot subdwarfs detected in UV all require mass transfer for their formation \citep{Heber2009ARA&A..47..211H}. None of the hot subdwarfs showed UV/IR excess flux, which indicates that their subdwarf component is the dominant flux emitter in the UV--IR range. 
Furthermore, the hotter companion candidates in three red clumps, four MS stars, two BSS and one post-HB star are also likely low luminosity sdAs/sdBs or ELMs according to their HR diagram position. If these are indeed hot subdwarfs or ELMs, then it means that all these are post mass transfer systems. 
Among the double SED-fitted MS stars, star33 is a known binary with a close companion. 
The MS stars which have accreted mass are known as blue lurkers \citep{Leiner2019ApJ...881...47L, Jadhav2022arXiv220703780J}. These are the MS equivalents of the BSSs. As such, they would be similar to BSSs in terms of formation scenarios (mergers or mass transfer) and detection techniques (high rotation, chemical peculiarities, post mass transfer companions).
Thus, star33 is a prime candidate for being a blue lurker. At the same time, the other three MS stars are also blue lurker candidates subject to confirmation of the companion's nature. If confirmed, this will make NGC 6791 one of the few clusters known to host blue lurkers. However, further observational evidence of rapid rotation and chemical alteration is necessary to confirm their status as blue lurkers.
We note that this cluster has several short-period binaries, suggestive of possible Case-A/Case-B mode of mass transfers, that can lead to binaries with very low mass stellar remnants (for example, star47). The UV images are able to detect relatively hotter systems, and there could be many more that are beyond the detection limit of UVIT. We note that such systems are found across the evolutionary stages from MS to post-HB.

The only detected RGB star showed UV excess when fitted with Kurucz models, which could later be explained by the updated Wein tail of UVBLUE models.
The other five giants (four red clumps and one post-HB) showed significant UV excess flux. NUV excess in RGBs has been known to be caused by rotation \citep{Dixon2020AJ....160...12D}. However, a hot compact companion can also give rise to UV excess. UV spectroscopy is required to confirm the exact source of the NUV flux from these RGBs.

Given the distance of NGC 6791, only the brightest WDs will be visible even in UV (all single subdwarfs detected presently are brighter than $\approx 6$ L$_{\odot}$). 
After cross-matching with \textit{Hubble Space Telescope} proper motion catalogue of NGC 6791 \citep{Libralato2022ApJ...934..150L}, we could identify one WD candidate (290.20235, 37.76336 at 21.83 mag in F814W) within the small footprint covered by the \textit{Hubble}. However, there were insufficient data points to parameterise it using its SED. There could be more such member WDs present, which could be discovered after deep multi-epoch wide-field imaging, such as the Legacy Survey of Space and Time (\citealt{Ivezic2019ApJ...873..111I}).

The colour of BSSs in the CMD has been linked with their origin. \citet{Ferraro2009Natur.462.1028F} claimed that bluer BSSs in globular cluster M 30 are formed via collisional mergers, while redder BSSs are formed via mass transfer. Similar double BSS sequences have been seen in other globular clusters (NGC 362: \citealt{Dalessandro2013ApJ...778..135D, Dattatrey2022arXiv221211302D}, NGC 2173: \citealt{Li2018ApJ...856...25L}).
Similarly, BSSs' UV excess has been linked to binarity \citep{Gosnell2015ApJ...814..163G, Sindhu2019ApJ...882...43S, Pandey2021MNRAS.507.2373P}.
In NGC 6791, the three BSSs with UV excess (indication of binarity) are generally redder than the rest of the BSSs of similar luminosity. This could be similar to the colour-dependent formation scenarios proposed by \citet{Ferraro2009Natur.462.1028F}. However, a bigger sample is needed to reach a statistically significant conclusion.

NGC 6791 has single sdBs and one sdA, along with possible low-luminous sdBs and ELMs as companions to stars in different evolutionary phases. All of these point to a very significant role that the binaries play in this dynamically old system, resulting in the formation of various types of stars due to differences in the masses of the binary members as well as their orbital properties. Therefore, This old open cluster is unique in possessing several features arising out of dynamics and binary systems.

\section{Summary \& Conclusion} \label{sec:conclusion}

UVIT detected 91 members of NGC 6791 in either far-UV or NUV filters, including MS, BSSs, hot subdwarfs, RGB and red clump stars. We provide SED-based parameters for 62 of the isolated sources.
Twenty of these sources showed UV excess flux, which is an indicator of a hot compact binary companion. We estimated approximate parameters for ten potential compact binary companions using SED analysis.

The total number of satisfactory binary fits is the same if we use a hybrid of UVBLUE+Kurucz models, but the number of sources with UV excess is more without UVBLUE. In general, we estimated higher temperatures for hotter companions when the
UVBLUE models were included.

NGC 6791 contains a significant population of single hot subdwarfs (four sdB and one sdA) and optically sub-luminous compact objects (ten hot subdwarf/ELMs). This is unique in open clusters and likely due to the dense and rich nature of NGC 6791. There are four blue lurker candidates in the cluster, which need to be confirmed using a combination of abundance studies, UV spectroscopy, rotational velocity, and RV/flux time series analysis. 
We note the presence of candidate post mass transfer systems across the evolutionary phases in this cluster, from MS, RGB, HB to post-HB, suggestive of a rich population of interacting binaries.

The cluster appears to show properties of a low-density globular cluster, along with some features seen in open clusters. This cluster is, therefore, a potential test bed, ideal for performing numerical simulations to understand the underlying processes regarding dynamics, binarity and stellar evolution.

\begin{acknowledgements}
    We thank the anonymous referee for constructive comments, which helped improve the methodology and discussion. VJ thanks the Alexander von Humboldt Foundation for their support. AS thanks for the support of the SERB power fellowship.
    RS thanks the National Academy of Sciences, India (NASI), Prayagraj, for the award of a NASI honorary Scientist position; the Alexander von Humboldt Foundation, Germany, for the award of Group linkage long-term research program between IIA, Bengaluru and European Southern Observatory, Munich, Germany, and the Director, IIA for providing institutional, infrastructural support during this work.
    UVIT project is a result of the collaboration between IIA, Bengaluru, IUCAA, Pune, TIFR, Mumbai, several centres of ISRO, and CSA. This publication uses \textsc{vosa}, developed under the Spanish Virtual Observatory project. 
\end{acknowledgements}

\bibliographystyle{aa} 
\bibliography{references} 

\end{document}